\documentstyle[aps,preprint,epsfig]{revtex}
\begin{document}
\tighten
\draft
\preprint{IFA-97/04}
\title{The Inverse Amplitude Method and Chiral Perturbation Theory to
two Loops}
\author{Torben Hannah\thanks{Electronic address: hannah@dfi.aau.dk}}
\address{Institute of Physics and Astronomy, Aarhus University,
DK-8000 Aarhus C, Denmark}
\maketitle
\begin{abstract}
The inverse amplitude method is analysed to two-loop order in the
chiral expansion in the case of $\pi\pi$ scattering and the pion form
factors. The analysis is mainly restricted to the elastic approximation
but the possible extension to the inelastic case is also discussed in
some detail. It is shown how the two-loop approach improves the
inverse amplitude method applied to one-loop order in the chiral
expansion. For both $\pi\pi$ scattering and the pion form factors, it
is in fact found that the inverse amplitude method to two-loop order
agrees remarkable well with the experimental data up to energies where
inelasticities become essential. At somewhat lower energies, the
two-loop approach compares well with the one-loop approximation, and in
the threshold region they both agree with chiral perturbation theory.
This suggests that the inverse amplitude method is indeed a rather
systematic way of improving chiral perturbation theory order by
order in the chiral expansion.
\end{abstract}

\section{INTRODUCTION}
\label{sec:intro}

The effective field theory of QCD, called chiral perturbation theory
(ChPT) \cite{ref:We79,ref:GL84,ref:Leut94}, has become a very
successful methodology for low-energy hadron physics. This effective
field theory is based upon the Goldstone character of the low-lying
pseudoscalar mesons and incorporates the appropriate symmetries of the
strong interaction. In this approach, one obtains a systematic
expansion in powers of external momentum and light quark masses, or
equivalently in the number of loops. However, in order to be
renormalized order by order, additional low-energy constants have to
be introduced at each order in the chiral expansion. These low-energy
constants are not fixed by chiral symmetry alone but have to be
determined phenomenologically.

At one-loop order in the chiral expansion only a small number of
low-energy constants have to be introduced \cite{ref:GL84}. These are
known rather accurately \cite{ref:GL84,ref:BCG94} and have been used
in order to relate different observables to each other. Therefore,
one-loop ChPT provides finite predictions for many different
low-energy observables in terms of only a very limited amount of
empirical input. Even though this has been remarkably successful, the
two-loop corrections are important in order to improve the accuracy
of the predictions. However, two-loop computations are very laborious
and have until now only been accomplished in a limited number of
cases. Furthermore, the total number of low-energy constants increases
dramatically at this order in the chiral expansion \cite{ref:FS96}.
Since an extensive phenomenological analysis of these new two-loop
low-energy constants is not available at present, they have so far
been evaluated either from the given process or from underlying
models.

Since ChPT provides a systematic perturbative expansion in powers of
external momentum and light quark masses, it will satisfy unitarity
perturbatively. This implies that the deviation from exact unitarity
at a given order provides an estimate of the energy range where the
truncation of the chiral expansion is applicable, i.e. the energy
range where higher order corrections can be neglected. Although these
higher order corrections could in principle be calculated and thereby
increase the energy range of the chiral expansion, ChPT is in
practice limited to the first few powers in the loop expansion. This
is due to the rapid increase in both computational effort and new
low-energy constants at each additional order. Thus, the unitarity
restriction will give a major limitation on the applicability of
ChPT. An important special case of this fact is in the presence of
resonances where higher order unitarity corrections become essential.

Therefore, in order to extend the range of applicability of ChPT,
unitarity will be of the utmost importance. A rather general way to
impose unitarity on the chiral expansion is by the inverse amplitude
method (IAM) \cite{ref:Tru88,ref:DHT90,ref:Han95,ref:Han96,ref:DP96},
which can be justified by the use of unitarity in a dispersive
approach. Since this method combines unitarity and dispersion
relations with the chiral expansion, it is likely that it will be more
powerful than ChPT alone. The IAM applied to one-loop order in the
chiral expansion has been extensively analysed
\cite{ref:Tru88,ref:DHT90,ref:Han95,ref:Han96,ref:DP96} and has indeed
proven to be a successful method to extend the range of applicability
of one-loop ChPT. However, the IAM is not restricted to the one-loop
approximation but can in general
be applied to any given order in the chiral
expansion. Therefore, it might well be that the IAM applied to a given
order can be improved by applying it to the next order. If this turns
out to be the case, the IAM could be considered as a systematic way
of improving ChPT order by order in the chiral expansion.

The IAM was originally applied to $\pi\pi$ scattering and the pion
form factors to one loop in ChPT \cite{ref:Tru88,ref:DHT90}. By now,
the full two-loop $\pi\pi$ scattering amplitude has been evaluated by
a dispersive analysis \cite{ref:Knec95} and a field theory calculation
\cite{ref:Bij96}, respectively. The pion form factors have also been
evaluated to two loops some time ago \cite{ref:GM90}. Therefore, it is
now possible to apply the IAM to two loops in the chiral expansion for
both $\pi\pi$ scattering and the pion form factors. In the present
paper, this extension of the IAM is analysed and compared to both the
one-loop case and one- and two-loop ChPT in order to study whether the
IAM is a systematic way of improving ChPT. Section \ref{sec:pipi} is
devoted to $\pi\pi$ scattering, Sec. \ref{sec:form} to the pion form
factors, and the conclusions are given in Sec. \ref{sec:con}.

\section{$\pi\pi$ SCATTERING}
\label{sec:pipi}

$\pi\pi$ scattering is of fundamental theoretical importance, both
because it only involves the self-interaction of the lightest
strongly interacting particle, and because it plays a significant
role for low-energy hadron phenomenology. The $\pi\pi$ scattering
amplitude can be expressed in terms of a single function
$A(s,t,u)$ as
\begin{equation}
\label{eq:pipiam}
T^{cd;ab}(s,t,u) = \delta^{ab}\delta^{cd}A(s,t,u) +
\delta^{ac}\delta^{bd}A(t,s,u) + \delta^{ad}\delta^{bc}A(u,t,s)
\end{equation}
where $s=(p^a+p^b)^2$, $t=(p^c-p^a)^2$, and $u=(p^d-p^a)^2$ are the
Mandelstam variables. The definite isospin amplitudes $T^I$ are given
in terms of the function $A(s,t,u)$ as
\begin{eqnarray}
\label{eq:isoam}
T^0(s,t,u) & = & 3A(s,t,u)+A(t,s,u)+A(u,t,s) ,\nonumber \\
T^1(s,t,u) & = & A(t,s,u)-A(u,t,s) ,\nonumber \\
T^2(s,t,u) & = & A(t,s,u)+A(u,t,s) ,
\end{eqnarray}
and the projection of these isospin amplitudes onto partial waves is
given by
\begin{equation}
\label{eq:partial}
t^I_l(s) = \frac{1}{64\pi}\int^1_{-1}d(\cos\theta )P_l(\cos\theta )
T^I(s,t,u) .
\end{equation}
In the elastic region $4M^2_{\pi}<s<16M^2_{\pi}$, unitarity implies
that the partial waves satisfy the following relation
\begin{equation}
\label{eq:pfuni}
{\rm Im}t^I_l(s) = \sigma (s)|t^I_l(s)|^2
\end{equation}
where $\sigma =\sqrt{(s-4M^2_{\pi})/s}$ is the phase-space factor.
As a consequence of Eq. (\ref{eq:pfuni}), the partial waves can be
parametrized in terms of real phase shifts
$\delta^I_l$ as
\begin{equation}
\label{eq:parpha}
t^I_l(s) = \frac{1}{\sigma (s)}\frac{1}{2i} \left\{
\exp[i2\delta^I_l(s)]-1 \right\} .
\end{equation}
The elastic unitarity relation (\ref{eq:pfuni}) is in practice useful
up to the $K\bar{K}$ threshold \cite{ref:Als71}. Above this energy the
extended unitarity relation becomes
\begin{equation}
\label{eq:epfuni}
{\rm Im}t^I_l(s) = \sigma (s)|t^I_l(s)|^2 + \sigma_{(K)}(s)
|t^I_{l(K)}(s)|^2
\end{equation}
where $t_{(K)}$ is the partial wave for the process
$\pi\pi\rightarrow K\bar{K}$ and $\sigma_{(K)}$ is the corresponding
phase-space factor. At still higher energies, other important
inelastic channels must be included and the unitarity relation has to
be generalized further. The inelastic channels can be included in
the parametrization (\ref{eq:parpha}) by adding an imaginary part to
the phase shifts.

\subsection{Chiral perturbation theory}
\label{subsec:ChPT}

In the framework of ChPT, the $\pi\pi$ scattering amplitude is
obtained as an expansion in powers of external momentum and light
quark masses
\begin{equation}
\label{eq:ChPTexam}
A(s,t,u) = A^{(0)}(s,t,u)+A^{(1)}(s,t,u)+A^{(2)}(s,t,u)+\cdots
\end{equation}
where $A^{(n)}$ is of $O(p^{2n+2})$. The leading order term $A^{(0)}$
corresponds to the current algebra result derived by Weinberg
\cite{ref:We66}, whereas $A^{(1)}$ is the one-loop correction
determined by Gasser and Leutwyler \cite{ref:GL84}. Recently, the
two-loop correction $A^{(2)}$ has also been evaluated by a dispersive
analysis \cite{ref:Knec95} and a field theory calculation
\cite{ref:Bij96}, respectively. These two approaches agree insofar as
the general structure of the amplitude is concerned. However, the
field theory calculation contains additional information regarding the
dependence of the chiral low-energy constants which is not obtained in
the dispersive analysis.

The full $\pi\pi$ scattering amplitude to two loops in ChPT is given
in Ref. \cite{ref:Bij96} where the contributions of $O(p^8)$ have
been omitted here. At one-loop order, there are four low-energy
constants $l^r_1$, $l^r_2$, $l^r_3$, and $l^r_4$, whereas six
additional low-energy constants $r^r_1$, $r^r_2$, $r^r_3$, $r^r_4$,
$r^r_5$, and $r^r_6$ have to be introduced at two-loop order. The
superscript $r$ indicates that these low-energy constants depend on
the renormalization scale $\mu$, whereas the full amplitude is scale
independent. The projection onto partial waves of the chiral $\pi\pi$
scattering amplitude (\ref{eq:ChPTexam}) gives the following expansion
\begin{equation}
\label{eq:ChPTpar}
t^I_l(s) = t^{I(0)}_l(s)+t^{I(1)}_l(s)+t^{I(2)}_l(s)+\cdots .
\end{equation}
These partial waves will satisfy the elastic unitarity relation
(\ref{eq:pfuni}) perturbatively, i.e. one has to two-loop order
\begin{eqnarray}
\label{eq:ppuni}
{\rm Im}t^{I(0)}_l(s) & = & 0 ,\nonumber \\
{\rm Im}t^{I(1)}_l(s) & = & \sigma (s){t^{I(0)}_l}^2(s) ,\nonumber \\
{\rm Im}t^{I(2)}_l(s) & = & \sigma (s)
2t^{I(0)}_l(s){\rm Re}t^{I(1)}_l(s) .
\end {eqnarray}
The first inelastic channel is the four pion intermediate state, which
is a three-loop effect of $O(p^8)$. Other inelastic channels with more
pions are of even higher order in the chiral expansion. This implies
that the perturbative elastic unitarity relations to two loops
will hold at least up to the $K\bar{K}$ threshold. In
fact, in the framework of SU(2) ChPT the relations (\ref{eq:ppuni})
are satisfied for all energies above threshold, whereas in SU(3) ChPT
they have to be generalized above the $K\bar{K}$ and
$\eta\eta$ thresholds. The generalization above the $K\bar{K}$
threshold is obtained from the perturbative interpretation of Eq.
(\ref{eq:epfuni})
\begin{eqnarray}
\label{eq:eppuni}
{\rm Im}t^{I(0)}_l(s) & = & 0 ,\nonumber \\
{\rm Im}t^{I(1)}_l(s) & = & \sigma (s){t^{I(0)}_l}^2(s)
+\sigma_{(K)}(s){t^{I(0)}_{l(K)}}^2(s) ,\nonumber \\
{\rm Im}t^{I(2)}_l(s) & = & \sigma (s)2t^{I(0)}_l(s){\rm Re}
t^{I(1)}_l(s)+\sigma_{(K)}(s)2t^{I(0)}_{l(K)}(s){\rm Re}
t^{I(1)}_{l(K)}(s) .
\end{eqnarray}
The further extension of these perturbative relations above the
$\eta\eta$ threshold is obtained in a straightforward manner.

\subsection{Inverse amplitude method}
\label{subsec:IAM}

The two-loop $\pi\pi$ scattering amplitude has hitherto only been
evaluated within the framework of SU(2) ChPT
\cite{ref:Knec95,ref:Bij96}. This implies that in the derivation of
the IAM applied to two loops in the chiral expansion, the elastic
approximation is used. However, the limitations of this approximation
and the possible extension to SU(3) ChPT will also be discussed in
some detail.

\subsubsection{Elastic approximation}
\label{subsubsec:pela}

From the fundamental principles of S matrix theory, it can be shown
that the $\pi\pi$ partial waves are analytic in the complex $s$ plane
with a right or unitarity cut for $4M^2_{\pi}<s<\infty$ and a left cut
for $-\infty <s<0$. Therefore, the $\pi\pi$ partial waves can be
expressed in terms of dispersion relations with a number of
subtractions to ensure the convergence of the integrals. The same is
also true for the inverse of the partial waves or more exactly for the
function $\Gamma ={t^{(0)}}^2/t$. Neglecting possible pole
contributions arising from zeros in the partial waves
\cite{ref:DHT90}, the four times subtracted dispersion relation for
the IAM is
\begin{equation}
\label{eq:dispinv}
\Gamma (s) = \Gamma_0+\Gamma_1s+\Gamma_2s^2+\Gamma_3s^3
+\frac{s^4}{\pi}\int_{4M^2_{\pi}}^{\infty}\frac{{\rm Im}\Gamma(s')ds'}
{s'^4(s'-s-i\epsilon )}
+\frac{s^4}{\pi}\int_{-\infty}^{0}\frac{{\rm Im}\Gamma(s')ds'}
{s'^4(s'-s-i\epsilon )} .
\end{equation}
Four subtractions are used in order to ensure that two-loop ChPT
satisfies a similar dispersion relation. On the right cut, the elastic
unitarity relation (\ref{eq:pfuni}) gives ${\rm Im}\Gamma
=-{t^{(0)}}^2{\rm Im}t/|t|^2=-\sigma {t^{(0)}}^2$, i.e. this part can
be computed exactly. The same is not true for the left cut, which,
however, can be approximated by means of two-loop ChPT. Expanding the
function $\Gamma$ to two loops in the chiral expansion $\Gamma
=t^{(0)}-t^{(1)}+{t^{(1)}}^2/t^{(0)}-t^{(2)}$, one finds that the left
cut can be approximated by ${\rm Im}\Gamma =-{\rm Im}t^{(1)}
+2{\rm Re}t^{(1)}{\rm Im}t^{(1)}/t^{(0)}-{\rm Im}t^{(2)}$. Finally,
the subtraction constants can also be evaluated to two-loop order in
the chiral expansion. Denoting these subtraction constants $a_0$,
$a_1$, $a_2$, and $a_3$, the following dispersion relation is obtained
for the IAM
\begin{eqnarray}
\label{eq:dispinv1}
\frac{{t^{I(0)}_l}^2(s)}{t^I_l(s)} & = & a_0+a_1s+a_2s^2+a_3s^3
-\frac{s^4}{\pi}\int_{4M^2_{\pi}}^{\infty}\frac{\sigma (s')
{t^{I(0)}_l}^2(s')ds'}{s'^4(s'-s-i\epsilon )} \nonumber \\ 
&&-\frac{s^4}{\pi}\int_{-\infty}^{0}
\frac{\left[ {\rm Im}t^{I(1)}_l(s')
-2{\rm Re}t^{I(1)}_l(s'){\rm Im}t^{I(1)}_l(s')/t^{I(0)}_l(s')
+{\rm Im}t^{I(2)}_l(s')\right] ds'}{s'^4(s'-s-i\epsilon )} .
\end{eqnarray}
This dispersion relation implies that the partial wave $t$ can be
obtained from the two-loop chiral expansion. However, the relation
between the IAM and two-loop ChPT can be simplified by writing down a
four-subtracted dispersion relation for the function $\Gamma^{(2)}
=t^{(0)}-t^{(1)}+{t^{(1)}}^2/t^{(0)}-t^{(2)}$. Since this is the
function $\Gamma$ expanded to two-loop order, the left cut and
subtraction constants in the dispersion relation for $\Gamma^{(2)}$
will be the same as in Eq. (\ref{eq:dispinv1}). For the right cut, the
perturbative elastic unitarity relations (\ref{eq:ppuni}) imply that
${\rm Im}\Gamma^{(2)}=-{\rm Im}t^{(1)}+2{\rm Re}t^{(1)}{\rm Im}t^{(1)}
/t^{(0)}-{\rm Im}t^{(2)}=-\sigma {t^{(0)}}^2+\sigma 2t^{(0)}{\rm Re}
t^{(1)}-\sigma 2t^{(0)}{\rm Re}t^{(1)}=-\sigma {t^{(0)}}^2$. Thus, the
following dispersion relation is obtained
\begin{eqnarray}
\label{eq:dispchi}
t^{I(0)}_l(s) & - & t^{I(1)}_l(s)+\frac{{t^{I(1)}_l}^2(s)}
{t^{I(0)}_l(s)}-t^{I(2)}_l(s) = a_0+a_1s+a_2s^2+a_3s^3
-\frac{s^4}{\pi}\int_{4M^2_{\pi}}^{\infty}\frac{\sigma (s')
{t^{I(0)}_l}^2(s')ds'}{s'^4(s'-s-i\epsilon )} \nonumber \\ 
& - & \frac{s^4}{\pi}\int_{-\infty}^{0}
\frac{\left[ {\rm Im}t^{I(1)}_l(s')
-2{\rm Re}t^{I(1)}_l(s'){\rm Im}t^{I(1)}_l(s')/t^{I(0)}_l(s')
+{\rm Im}t^{I(2)}_l(s')\right] ds'}{s'^4(s'-s-i\epsilon )} ,
\end{eqnarray}
which is exactly the same as in Eq. (\ref{eq:dispinv1}). Therefore,
the IAM to two loops in the chiral expansion gives the partial wave
$t$ in the simple form
\begin{equation}
\label{eq:Pade2}
t^I_l(s) = \frac{t^{I(0)}_l(s)}{1-t^{I(1)}_l(s)/t^{I(0)}_l(s)
+{t^{I(1)}_l}^2(s)/{t^{I(0)}_l}^2(s)-t^{I(2)}_l(s)/t^{I(0)}_l(s)} .
\end{equation}
Since this is formally equivalent to the [0,2] Pad\'{e} approximant
applied on ChPT, the expression (\ref{eq:Pade2}) will coincide with
the chiral expansion up to two-loop order. However, in deriving
(\ref{eq:Pade2}) two-loop ChPT has only been used for the left cut
and the subtraction constants, whereas the important unitarity cut was
computed exactly. This is in contrast to ChPT where the unitarity cut
was only computed perturbatively. As a consequence of this, the [0,2]
Pad\'{e} approximant (\ref{eq:Pade2}) satisfies the elastic unitarity
relation (\ref{eq:pfuni}) exactly and not only perturbatively as in
ChPT.

The IAM can in general be applied to any given order in the chiral
expansion, but has up to now almost entirely been restricted to the
one-loop approximation
\cite{ref:Tru88,ref:DHT90,ref:Han95,ref:Han96,ref:DP96}. In this case,
the IAM gives the partial waves in the form
\begin{equation}
\label{eq:Pade1}
t^I_l(s) = \frac{t^{I(0)}_l(s)}{1-t^{I(1)}_l(s)/t^{I(0)}_l(s)} ,
\end{equation}
which is formally equivalent to the [0,1] Pad\'{e} approximant. This
form also satisfies unitarity exactly and coincides with the chiral
expansion up to one-loop order. It is derived in the same way
as described above, the only difference being that the left cut and
subtraction constants in Eq. (\ref{eq:dispinv1}) are now approximated
only to one-loop order in the chiral expansion. This shows how the IAM
to two loops improves the IAM to one loop, even though the dominantly
unitarity cut is computed exactly in both cases.

The above discussion rests upon the assumption that the leading order
term $t^{(0)}$ does not vanish. In fact, this is only true for the
lowest partial waves with $l\leq 1$, whereas the higher partial waves
start at one-loop order in the chiral expansion. In these cases, the
IAM must be slightly modified by writing down a dispersion relation
for the function ${t^{(1)}}^2/t$. Since perturbative unitarity implies
that $t^{(1)}$ is real on the right cut for these higher partial
waves, the elastic unitarity relation (\ref{eq:pfuni}) gives
${\rm Im}({t^{(1)}}^2/t)=-{t^{(1)}}^2{\rm Im}t/|t|^2=-\sigma
{t^{(1)}}^2$. Hence, the right cut can also in these cases be computed
exactly, whereas the left cut and subtraction constants can be
approximated by means of ChPT as above. However, in order to relate
the dispersion relation for the IAM to a corresponding one for ChPT,
one has to go to three loops in the chiral expansion. This is due to
the fact that for the higher partial waves the imaginary part on the
right cut starts at $O(p^8)$ in ChPT. Should such a three-loop
calculation be undertaken in the future, the IAM could also in these
cases be directly related to ChPT in the same way as discussed above.

\subsubsection{Inclusion of inelasticities}
\label{subsubsec:pinela}

It would be desirable if the IAM could be applied beyond
the elastic approximation. The first important inelastic channel in
$\pi\pi$ scattering opens up at the $K\bar{K}$ threshold and is due to
the $K\bar{K}$ intermediate state. At a slightly higher energy the
$\eta\eta$ inelastic channel will also appear, but this is a rather
unimportant effect. Only at still higher energies will the four-pion
intermediate state begin to be significant, together with other
inelastic channels. The $K\bar{K}$ and $\eta\eta$ intermediate states
are already present at one-loop order in SU(3) ChPT \cite{ref:BKM91},
and will therefore also appear in an extension of the present two-loop
calculation to the SU(3) case. Thus, it might be that the IAM applied
to two loops in the SU(3) chiral expansion could include these
inelasticities. In order to try to include the four-pion intermediate
state, one would need to apply the IAM to three loops since this
effect is of $O(p^8)$ in the chiral expansion.

Therefore, the following discussion is restricted to the $K\bar{K}$ and
$\eta\eta$ inelastic channels. In fact, it will be limited to the
important $K\bar{K}$ case since the extension to the
$K\bar{K}/\eta\eta$ case is rather obvious. To include the $K\bar{K}$
inelastic channel, the extended unitarity relation (\ref{eq:epfuni})
has to be used above the $K\bar{K}$ threshold in computing the right
cut in Eq. (\ref{eq:dispinv1}), whereas the other parts are computed
in exactly the same way as before. Thus, since the relation
(\ref{eq:epfuni}) gives ${\rm Im}\Gamma =-{t^{(0)}}^2{\rm Im}t/|t|^2
=-\sigma {t^{(0)}}^2-\sigma_{(K)}{t^{(0)}}^2|t_{(K)}|^2/|t|^2$, this
expression has to be used above the $K\bar{K}$ threshold in Eq.
(\ref{eq:dispinv1}). However, in this case it is not possible to
compute the inelastic part $|t_{(K)}|^2/|t|^2$ exactly, so this part
must be evaluated to two loops in the SU(3) chiral expansion
\begin{eqnarray}
\label{eq:inelaex}
{\rm Im}\Gamma (s) & = & -\sigma (s){t^{I(0)}_l}^2(s)-\sigma_{(K)}(s)
{t^{I(0)}_l}^2(s)\frac{|t^I_{l(K)}(s)|^2}{|t^I_l(s)|^2} \nonumber \\
&\approx & -\sigma (s){t^{I(0)}_l}^2(s)-\sigma_{(K)}(s)
{t^{I(0)}_{l(K)}}^2(s) \nonumber \\
&&-\sigma_{(K)}(s)2{t^{I(0)}_{l(K)}}^2(s)
\left[ \frac{{\rm Re}t^{I(1)}_{l(K)}(s)}{t^{I(0)}_{l(K)}(s)}
-\frac{{\rm Re}t^{I(1)}_l(s)}{t^{I(0)}_l(s)} \right] .
\end{eqnarray}
In order to simplify the relation between the IAM and ChPT for the
SU(3) case, the right cut in Eq. (\ref{eq:dispchi}) has to be
evaluated with the inclusion of the $K\bar{K}$ intermediate state.
From the extended perturbative unitarity relations
(\ref{eq:eppuni}), one finds that ${\rm Im}\Gamma^{(2)}$ is given by
an expression exactly similar to Eq. (\ref{eq:inelaex}). Thus, the
inelastic case also gives a result of the form (\ref{eq:Pade2}), the
only difference being that in this case the chiral SU(3) expansion is
used. In a similar way, by expanding the inelastic part
$|t_{(K)}|^2/|t|^2$ to one-loop order, the IAM gives a result of the
form (\ref{eq:Pade1}) with ChPT evaluated within SU(3).

In the presence of the $K\bar{K}$ inelasticity, the unitarity cut
cannot be computed exactly, but one has to expand the inelastic part to
a given chiral order. Since the inelastic part starts at the
$K\bar{K}$ threshold, the one- and two-loop chiral expansions would
seem to lead to rather unreliable approximations.
However, it is important to realize that for the inelastic IAM the
exact extended unitarity relation has been applied before the
inelastic part is approximated by the chiral expansion. Furthermore,
if one compare the expansion of the inelastic part for the IAM
(\ref{eq:inelaex}) with the corresponding expansion for ChPT
(\ref{eq:eppuni}), the two-loop correction in Eq. (\ref{eq:inelaex})
is likely to be more suppressed compared to the two-loop correction in
Eq. (\ref{eq:eppuni}). Therefore, one might expect that the expansion
of the inelastic quantity $|t_{(K)}|^2/|t|^2$ converges rather rapidly,
i.e. the one- and two-loop expansion could be used over a large energy
range. Indeed, as will be shown in the case of the pion form factors,
the expansion of a similar quantity works very well up to rather high
energies.

In order to evaluate how well the inelastic IAM agrees with
the extended unitarity relation (\ref{eq:epfuni}), one needs to
determine the $t_{(K)}$ partial waves. To do this, the IAM could be
applied to one- and two-loop order for these partial waves. In this
way one finds that the inelastic IAM satisfies Eq. (\ref{eq:epfuni})
exactly, provided that the quantity $|t_{(K)}|^2/|t|^2$ is determined
by the leading order term. Thus, the IAM is a rather consistent
approach, but there is of course no way to know a priori to what extent
this method can reproduce the inelasticities.

\subsection{Comparison with experiment}
\label{subsec:pcom}

The chiral expansion depends on a number of low-energy constants not
fixed by chiral symmetry alone. These have to be determined before
ChPT and the IAM can be compared with experimental data. In the
present paper, this comparison is limited to the lowest $\pi\pi$
partial waves with $l\leq 1$, since they are the best known
experimentally.

\subsubsection{The low-energy constants}
\label{subsubsec:lec}

Since the chiral low-energy constants depend on the renormalization
scale $\mu$, this scale has to be chosen before the low-energy
constants can be fixed. This scale will be set at the mass of the
$\rho$(770) resonance, i.e. $\mu =M_{\rho}=770$ MeV. In addition,
throughout this paper the
values $F_{\pi}=92.4$ MeV and $M_{\pi}=139.6$ MeV are used for the
pion decay constant and pion mass, respectively.

The $\pi\pi$ partial waves depend on four low-energy constants
$l^r_1$, $l^r_2$, $l^r_3$, and $l^r_4$ to one-loop order in the chiral
expansion. These have been determined phenomenologically rather
accurately in one-loop ChPT \cite{ref:GL84,ref:BCG94} with results
displayed in the first column in Table \ref{Table1}. For ChPT to two
loops, these low-energy constants should in principle be re-evaluated
taking into account the two-loop corrections. However, since
a thorough reevaluation seems
rather out of reach at the moment, the one-loop values of $l^r_1$,
$l^r_2$, $l^r_3$, and $l^r_4$ are also used in two-loop ChPT. The
additional low-energy constants to two loops $r^r_1$, $r^r_2$,
$r^r_3$, $r^r_4$, $r^r_5$, and $r^r_6$ may be estimated with the
assumption that they are dominated by resonance contributions at the
scale of the resonance, an assumption that is well satisfied in
one-loop ChPT \cite{ref:GL84,ref:Eck89}. In particular, including
contributions from vector and scalar exchange together with terms of
$O(p^6)$ coming from SU(3) one-loop ChPT \cite{ref:Bij96}, one obtains
the values for these two-loop low-energy constants given in the second
column in Table \ref{Table1} \cite{ref:Bij96a}.

The IAM to a given order depends on the same low-energy constants as
ChPT to the same order. However, the phenomenological determination of
these low-energy constants by the IAM do not necessarily coincide
precisely with the values obtained in ChPT. This is due to the fact
that the IAM contains higher order unitarity corrections, which will
effect the phenomenological determination of the low-energy constants.
Some of the low-energy constants in the IAM to one and two loops may
be determined phenomenologically by fitting the experimental $\pi\pi$
phase shifts. In this fitting, the data obtained from $K_{l4}$ decays
\cite{ref:Ros77} are used together with the data from high statistics
medium energy experiments
\cite{ref:Proto73,ref:Hyams73,ref:EM74,ref:Losty74,ref:Hoog77}. For a
recent discussion of the experimental
$\pi\pi$ scattering data see Ref. \cite{ref:MP95}.

The low-energy constants $l^r_1$ and $l^r_2$ in the
IAM to one loop may be determined in this way, whereas the fit will be
rather insensitive to the precise values of the other low-energy
constants $l^r_3$ and $l^r_4$. For these latter low-energy constants,
the values obtained in one-loop ChPT are also used in the IAM to one
loop. In the case of $l^r_1$ and $l^r_2$, one is
used to fix the $\rho$(770) resonance defined to be at
$\delta^1_1(M_{\rho})=\pi /2$, whereas the other is determined by
fitting the $\delta^0_0$ and $\delta^2_0$ experimental phase shifts
for $\sqrt{s}\leq 0.7$ GeV. This will give an acceptable
fit with values of $l^r_1$ and $l^r_2$ given in the third
column in Table \ref{Table1}, whereas if one increase the energy range
to $\sqrt{s}\leq 0.9$ GeV, the fit will not be acceptable.

However, the IAM to two loops might
improve the agreement with the experimental phase shifts up to
$\sqrt{s}=0.9$ GeV. In this case the fit is mainly sensitive to
$l^r_1$, $l^r_2$, $r^r_5$, and $r^r_6$, so the other low-energy
constants will be given by the same values as in two-loop ChPT.
With the use of the $\rho$(770) resonance
to fix $r^r_6$ and varying $l^r_1$,
$l^r_2$, and $r^r_5$ around the values used in two-loop ChPT, these
low-energy constants are determined by fitting the $\delta^0_0$ and
$\delta^2_0$ experimental phase shifts up to $\sqrt{s}=0.9$ GeV. In
this case one obtains an acceptable fit with values of the
low-energy constants given in the fourth column in Table \ref{Table1}.
There is no point in extending this fit further up in energy since
inelastic effects will become essential above approximately 0.9 GeV.

The low-energy constants in Table \ref{Table1} have not been assigned
any error bars since it is difficult to estimate these error bars
taking into account the higher order corrections. In addition, it is
not possible to assign reliable error bars on the two-loop low-energy
constants estimated from resonance exchange. Nevertheless, from Table
\ref{Table1} it is observed that the values of the low-energy
constants in ChPT and the IAM to both one and two loops are rather
consistent with each other. However, some variations are indeed
to be expected due to the different treatments of the higher order
corrections.

\subsubsection{Threshold parameters and phase shifts}
\label{subsubsec:pthres}

Close to threshold, the $\pi\pi$ partial waves can be parametrized in
terms of scattering lengths $a^I_l$ and slope parameters $b^I_l$
\cite{ref:Nagels79}. In Table \ref{Table2}, the values obtained from
ChPT and the IAM are compared to the experimental information for
these threshold parameters. It is observed that one-loop ChPT gives
relatively large corrections to the leading order current algebra
results, whereas the additional two-loop ChPT corrections are somewhat
smaller. However, the difference between one and two-loop ChPT is in
general larger than the corresponding difference between the IAM to
one and two loops, which supports the expectation that the IAM to one
loop already includes the most important unitarity corrections. In
fact, the IAM to both one and two loops compares remarkably well with
two-loop ChPT for the $I=0$ S wave, where the unitarity corrections
are important even at low energies. For the other channels, where the
higher order unitarity corrections are less important at low energies,
the IAM compares in general rather well with both one and two-loop
ChPT.

Regarding the comparison with the experimental information, both
the IAM and ChPT to one and two loops are quite consistent with the
data. Unfortunately, the precision of these data does not allow one
to distinguish the higher order corrections from the one-loop
contribution, but this might be improved with the forthcoming
accurately determination of $a^0_0-a^2_0$ from the measurement
of the lifetime of the $\pi^+\pi^-$ atom \cite{ref:Ade95}.

\begin{figure}
\centering{\epsfig{file=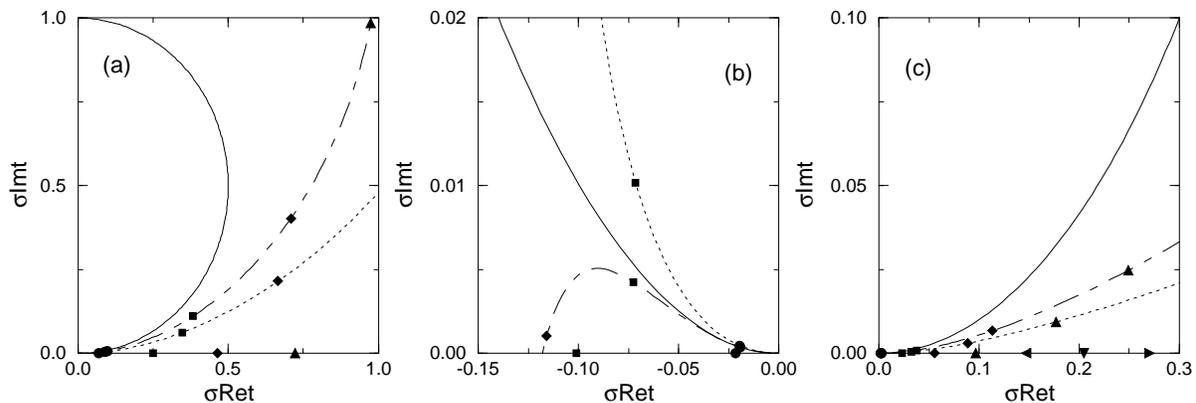,angle=-90}}
\caption{Argand diagrams for the ChPT partial waves $t^0_0$ (a),
$t^2_0$ (b), and $t^1_1$ (c), respectively. The solid line is the
unitarity circle, the dashed-dotted line two-loop ChPT, and the dotted
line one-loop ChPT. The leading order current algebra result lies on
the real axis of the diagrams.
The corresponding energies are displayed
on the curves for 0.3 GeV (circles), 0.4 GeV (squares), 0.5 GeV
(diamonds), 0.6 GeV (up triangles), 0.7 GeV (left triangles), 0.8 GeV
(down triangles), and 0.9 GeV (right triangles).}
\label{Fig1}
\end{figure}

Even though the chiral expansion only satisfies unitarity
perturbatively, ChPT is not in general restricted to the threshold
region. However, the deviation from exact unitarity will give an
estimate of the actual energy range where the given truncation of the
chiral expansion is applicable. This deviation is shown in the Argand
diagrams in Fig. \ref{Fig1} for the three lowest partial waves. This
shows that unitarity is indeed improved at low energies at each
additional order in the chiral expansion. On the other hand, two-loop
ChPT begins to deviate significantly from exact unitarity above
approximately 0.5 GeV for all three partial waves. Therefore, two-loop
ChPT should only be trusted up to about this energy, whereas the
energy range for one-loop ChPT is somewhat smaller.

Despite the fact that ChPT only satisfies unitarity perturbatively,
it is possible to define a chiral expansion for the phase shifts
$\delta^I_l$ \cite{ref:Knec95,ref:GM91}. In Fig. \ref{Fig2}, the phase
shift difference $\delta^0_0-\delta^1_1$ given by the IAM and ChPT is
compared to the experimental data obtained from $K_{l4}$ decays
\cite{ref:Ros77}. The IAM to one and two loops gives very similar
results in the displayed energy region. Furthermore, since these
results compare very well to both two-loop ChPT and the central
experimental data, the higher order unitarity corrections are indeed
very well approximated by the IAM.
As for one-loop ChPT, this appears to
be systematically below the central experimental data, although it is
still consistent with these data within the error bars. At DA$\Phi$NE,
it is expected that these error bars can be significantly reduced from
new high statistics measurements of the $K_{l4}$ decays
\cite{ref:BF95}. From this, it will also be possible to determine
accurately some of the low-energy constants in the IAM from
independent observables \cite{ref:Han95} and thereby test the
consistency of the IAM in great detail.

In Fig. \ref{Fig3}, the phase shift $\delta^0_0$ is shown below
$\sqrt{s}=1$ GeV. It is observed that the IAM to two loops agrees
rather well with the data up to approximately 0.9 GeV, whereas the IAM
to one loop only describes the data well up to about 0.7 GeV. For one
and two-loop ChPT, both begin to disagree with the data somewhat above
0.5 GeV, which is consistent with
the unitarity requirement. In fact, the IAM to two loops describes the
data rather well up to energies where the elastic approximation will no
longer be applicable. Therefore, before one tries to apply this
method to even higher energies, the $K\bar{K}$ inelasticity must be
included in the IAM as discussed previously. However, whether this
inelastic IAM will actually be able to reproduce the $f_0$(980)
resonance, which is closely related to the $K\bar{K}$ inelastic channel,
has to await further investigations.

In Fig. \ref{Fig4}, the phase shift $\delta^2_0$ is shown below
$\sqrt{s}=1.4$ GeV. In this case, the IAM to two loops also seems to
improve the IAM to one loop at higher energies, even though the data
are not very conclusive. In addition, the IAM also in this case
successfully extends the range of applicability of ChPT. In this
channel, the first important inelastic effect is due to the four
pion intermediate state, which should begin to be significant somewhat
above 1.2 GeV. Thus, the IAM to two loops describes the data well up
to energies where this inelasticity should be included. It also
seems that two-loop ChPT is rather consistent with the data all the
way up to about 1 GeV. However, this is in conflict with the unitarity
requirement, which shows that two-loop ChPT can only be trusted up
to approximately 0.5 GeV.

Finally, in Fig. \ref{Fig5} the phase shift $\delta^1_1$ is shown
below $\sqrt{s}=1.4$ GeV. This channel is dominated by the $\rho$(770)
resonance, which was used in the determination of the low-energy
constants for the IAM. However, it is observed that the IAM is not
only consistent with the experimental data close to the resonance
mass, but also for energies rather far away from this resonance.
Furthermore, for energies somewhat above 1 GeV the IAM to two loops
improves the IAM to one loop, although inelastic effects should begin
to be significant at these energies. As for ChPT, this only describes
the low-energy tail of the $\rho$(770) resonance.

It has recently been shown that the IAM gives the correct analytical
structure in the complex $s$ plane \cite{ref:DP96}. Of course, this
method produces the appropriate cuts on the first Riemann sheet, since
these cuts are already present in ChPT. However, contrary to ChPT, the
IAM can also produce poles on the second Riemann sheet, where
resonances are related to such poles in the vicinity of the real axis.
Indeed, the IAM gives the correct pole structure associated with the
$\rho$(770) resonance. Furthermore, consistent with recent
phenomenological analysis \cite{ref:PDG96}, this method also gives a
pole in the $I=0$ S wave responsible for the strong final-state
interaction in this channel. In fact, these poles have only been
obtained for the IAM to one loop \cite{ref:DP96} but the same should
also be true in general. Since this analytical structure is not
trivial at all, this strongly supports the applicability of the IAM.

\section{PION FORM FACTORS}
\label{sec:form}

The pion vector form factor ($F_V$) provides important information
about the internal structure of the pion. Another quantity of interest
is the pion scalar form factor ($F_S$) defined by the matrix element
of the quark density. These form factors depend on one kinematical
variable $s$, giving by the square of the four-momentum transfer. They
are analytical functions in the complex $s$ plane with a unitarity cut
starting at the $\pi\pi$ threshold. The form factors will only be
considered in the isospin limit, and the scalar form factor will be
normalized according to $F_S(0)=1$. Thus, they may be expanded around
$s=0$ as
\begin{equation}
\label{eq:Fex}
F(s) = 1+\mbox{$\frac{1}{6}$}\langle r^2\rangle s+cs^2+\cdots
\end{equation}
where $F$ is a generic symbol for the vector and scalar form factors.
In the elastic region, one has the following unitarity relation
\begin{equation}
\label{eq:ffuni}
{\rm Im}F(s) = \sigma (s)F^{\ast}(s)t^I_l(s) ,
\end{equation}
which implies that the phase of $F$ will coincide with the $\pi\pi$
phase shift $\delta^I_l$ in accordance with Watson's final-state
theorem \cite{ref:Wat54}. More precisely, in the elastic region the
phase of $F_V$ will coincide with the $\pi\pi$ $I=1$ P phase shift
$\delta^1_1$, whereas the phase of $F_S$ will be given by the $I=0$ S
phase shift $\delta^0_0$. Above the $K\bar{K}$ threshold, the
elastic unitarity relation (\ref{eq:ffuni}) has to be modified due
to the important $K\bar{K}$ inelasticity. This extended unitarity
relation can be written as
\begin{equation}
\label{eq:fefuni}
{\rm Im}F(s) = \sigma (s)F^{\ast}(s)t^I_l(s)+\sigma_{(K)}
F^{\ast}_{(K)}(s)t^I_{l(K)}(s)
\end{equation}
where $F_{(K)}$ is the properly normalized kaon vector or scalar form
factor. At still higher energies, other important inelastic channels
must be included and the unitarity relation has to be generalized
further.

\subsection{Chiral perturbation theory}
\label{subsec:FChPT}

The chiral expansion of the pion vector and scalar form factors can
be written as
\begin{equation}
\label{eq:Cformex}
F(s) = F^{(0)}(s)+F^{(1)}(s)+F^{(2)}(s)+\cdots
\end{equation}
where $F^{(n)}$ is of $O(p^{2n})$. Since the scalar form factor is
normalized according to $F_S(0)=1$, one has for the leading order term
$F^{(0)}=1$ for both form factors. Regarding the one-loop correction
$F^{(1)}$, this has been determined by Gasser and Leutwyler both in
SU(2) ChPT \cite{ref:GL84} and in the SU(3) case \cite{ref:GL85}.

For the two-loop correction $F^{(2)}$, this part has been evaluated by
a dispersive analysis \cite{ref:GM90} in the elastic SU(2)
approximation. This dispersive approach is based upon the two-loop
perturbative interpretation of the elastic unitarity relation Eq.
(\ref{eq:ffuni})
\begin{eqnarray}
\label{eq:fpuni}
{\rm Im}F^{(0)}(s) & = & 0 , \nonumber \\
{\rm Im}F^{(1)}(s) & = & \sigma (s)t^{I(0)}_l(s) , \nonumber \\
{\rm Im}F^{(2)}(s) & = & \sigma (s) \left[ {\rm Re}F^{(1)}(s)
t^{I(0)}_l(s)+{\rm Re}t^{I(1)}_l(s) \right] .
\end{eqnarray}
These relations are satisfied for all energies above the $\pi\pi$
threshold in the SU(2) approximation. Therefore, in this case
the full two-loop pion form factors may be expressed in terms of a
dispersion relation as \cite{ref:GM90}
\begin{equation}
\label{eq:Cform}
F(s) = 1+\mbox{$\frac{1}{6}$}\langle r^2\rangle s+cs^2
+\frac{s^3}{\pi}\int^{\infty}_{4M^2_{\pi}}\frac{\sigma (s') \left\{
t^{I(0)}_l(s')\left[ 1+{\rm Re}F^{(1)}(s')\right] +{\rm
Re}t^{I(1)}_l(s')\right\} ds'}{s'^3(s'-s-i\epsilon )} .
\end{equation}
For the vector form factor, the subtraction constants $\langle
r^2\rangle$ and $c$ can be written as
\begin{eqnarray}
\label{eq:Vr2c}
\langle r^2\rangle_V & = & \frac{1}{16\pi^2F^2_{\pi}}\left[
(\bar{l}_6-1)+\frac{\bar{f}_1M^2_{\pi}}{16\pi^2F^2_{\pi}} \right]
,\nonumber \\
c_V & = & \frac{1}{16\pi^2F^2_{\pi}}\left[ \frac{1}{60M^2_{\pi}}
+\frac{\bar{f}_2}{16\pi^2F^2_{\pi}}\right] ,
\end{eqnarray}
whereas in the case of the scalar form factor, they may be
expressed in the following way:
\begin{eqnarray}
\label{eq:Sr2c}
\langle r^2\rangle_S & = & \frac{3}{8\pi^2F^2_{\pi}}\left[
(\bar{l}_4-\mbox{$\frac{13}{12}$})+\frac{\bar{d}_1M^2_{\pi}}
{16\pi^2F^2_{\pi}} \right] , \nonumber \\
c_S & = & \frac{1}{16\pi^2F^2_{\pi}}\left[ \frac{19}{120M^2_{\pi}}
+\frac{\bar{d}_2}{16\pi^2F^2_{\pi}}\right] .
\end{eqnarray}
The subtraction constants are given in terms of
the scale-independent one-loop
low-energy constants $\bar{l}_4$ and $\bar{l}_6$ and
the additional two-loop coupling constants $\bar{f}_1$, $\bar{f}_2$,
$\bar{d}_1$, and $\bar{d}_2$. These coupling constants could in the
future be expressed in terms of chiral logs and two-loop renormalized
chiral low-energy constants.

The dispersive analysis is not restricted to the elastic SU(2)
approximation, but may be generalized to the inelastic case as well.
The important $K\bar{K}$ inelasticity can be included in this approach
by considering the two-loop perturbative extended unitarity relations
\begin{eqnarray}
\label{eq:fepuni}
{\rm Im}F^{(0)}(s) & = & 0 , \nonumber \\
{\rm Im}F^{(1)}(s) & = & \sigma (s)t^{I(0)}_l(s)+\sigma_{(K)}(s)
F^{(0)}_{(K)}(s)t^{I(0)}_{l(K)}(s) , \nonumber \\
{\rm Im}F^{(2)}(s) & = & \sigma (s) \left[ {\rm Re}F^{(1)}(s)
t^{I(0)}_l(s)+{\rm Re}t^{I(1)}_l(s) \right] \nonumber \\
&&+\sigma_{(K)}(s) \left[ {\rm Re}F^{(1)}_{(K)}(s)t^{I(0)}_{l(K)}(s)
+F^{(0)}_{(K)}(s){\rm Re}t^{I(1)}_{l(K)}(s) \right] .
\end{eqnarray}
In the inelastic case, these expressions have to be used above the
$K\bar{K}$ threshold in the dispersion relation Eq. (\ref{eq:Cform}).
The other inelastic channels may also be included in the dispersive
approach, but these will only be important somewhat above the
$K\bar{K}$ threshold.

\subsection{Inverse amplitude method}
\label{subsec:FIAM}

Also for the form factors, the elastic approximation will be used
in the derivation of the IAM applied to two loops in the chiral
expansion. However, some short remarks on the possible inclusion
of inelasticities will also be given in this case.

\subsubsection{Elastic approximation}
\label{subsubsec:fela}

The starting point for the IAM applied to the pion form factors is to
write down a dispersion relation for the inverse of the form factor
$\Gamma =1/F$. In this dispersion relation, the elastic unitarity
relation (\ref{eq:ffuni}) gives ${\rm Im}\Gamma =-{\rm Im}F/|F|^2
=-\sigma t/F$. Unfortunately, this part cannot be computed
exactly, as was the case for $\pi\pi$ scattering, but it may be
expanded to two-loop order in the chiral expansion ${\rm Im}\Gamma
=-\sigma [t^{(0)}(1-{\rm Re}F^{(1)})+{\rm Re}t^{(1)}]$. The
subtraction constants can also be evaluated by expanding the
function $\Gamma$ to two-loop order as $\Gamma =1-F^{(1)}
+{F^{(1)}}^2-F^{(2)}$. Thus, neglecting any pole contribution arising
from possible zeros in the form factors \cite{ref:Tru88,ref:DHT90},
the following dispersion relation is obtained for the IAM
\begin{equation}
\label{eq:fdisp1}
\frac{1}{F(s)} = 1+a_1s+a_2s^2-\frac{s^3}{\pi}
\int^{\infty}_{4M^2_{\pi}}\frac{\sigma (s')\left\{ t^{I(0)}_l(s')
\left[ 1-{\rm Re}F^{(1)}(s')\right] +{\rm Re}t^{I(1)}_l(s')\right\}
ds'}{s'^3(s'-s-i\epsilon )} .
\end{equation}
In order to simplify the connection between the IAM and two-loop ChPT,
one can write down a dispersion relation for the function
$\Gamma^{(2)}=1-F^{(1)}+{F^{(1)}}^2-F^{(2)}$. Since this is the
function $\Gamma$ expanded to two-loop order in the chiral expansion,
the subtraction constants in the dispersion relation for
$\Gamma^{(2)}$ will be the same as in Eq. (\ref{eq:fdisp1}).
Furthermore, the perturbative elastic unitarity relations
(\ref{eq:fpuni}) give ${\rm Im}\Gamma^{(2)}=-{\rm Im}F^{(1)}
+2{\rm Re}F^{(1)}{\rm Im}F^{(1)}-{\rm Im}F^{(2)}=-\sigma t^{(0)}
+\sigma 2{\rm Re}F^{(1)}t^{(0)}-\sigma {\rm Re}F^{(1)}t^{(0)}
-\sigma {\rm Re}t^{(1)}=-\sigma [t^{(0)}(1-{\rm Re}F^{(1)})
+{\rm Re}t^{(1)}]$, so the following dispersion relation is
obtained for ChPT
\begin{eqnarray}
\label{eq:fdisp2}
1 - && F^{(1)}(s)+{F^{(1)}}^2(s)-F^{(2)}(s) = 1+a_1s+a_2s^2
\nonumber \\
-\frac{s^3}{\pi}\int^{\infty}_{4M^2_{\pi}} && \frac{\sigma (s')
\left\{ t^{I(0)}_l(s')\left[ 1-{\rm Re}F^{(1)}(s')\right]
+{\rm Re}t^{I(1)}_l(s')\right\} ds'}{s'^3(s'-s-i\epsilon )} .
\end{eqnarray}
Since this is identical to Eq. (\ref{eq:fdisp1}), the IAM to two-loop
order in the chiral expansion can be written as
\begin{equation}
\label{eq:FPade2}
F(s) = \frac{1}{1-F^{(1)}(s)+{F^{(1)}}^2(s)-F^{(2)}(s)} .
\end{equation}
Even though the unitarity cut in the IAM for the pion form factors
cannot be computed exactly, it is still expected that this method is
superior to the truncation of the chiral expansion. This is based upon
the fact that the expansion of $t/F$ used in the IAM converges
significantly faster than the corresponding expansion of
$F^{\ast}t$ used in ChPT. Actually, even the one-loop expansion
$t/F=t^{(0)}$ is a rather good approximation over a
relatively large energy region \cite{ref:Tru88}. With this
approximation and the $\pi\pi$ partial waves given by Eq.
(\ref{eq:Pade2}), the IAM to two loops will satisfy the elastic
unitarity relation (\ref{eq:ffuni}) exactly. Hence, the IAM is
a rather consistent way of imposing the unitarity restriction on the
chiral expansion.

The IAM was originally applied to the pion form factors in the
one-loop approximation \cite{ref:Tru88}, which gives the result
\begin{equation}
\label{eq:FPade1}
F(s) = \frac{1}{1-F^{(1)}(s)} .
\end{equation}
With the $\pi\pi$ partial waves given by Eq. (\ref{eq:Pade1}), this
form will also satisfy the elastic unitarity relation (\ref{eq:ffuni})
exactly, provided that the one-loop approximation $t/F=t^{(0)}$
is satisfied. However, in this case $t/F$ and the subtraction
constants are only expanded to one-loop order in the chiral expansion.
Thus, even though the one-loop expansion of $t/F$ is a rather good
approximation, it is clear that the IAM to two loops will improve this
approximation.

\subsubsection{Inclusion of inelasticities}
\label{subsubsec:finela}

The first important inelastic channel is due to the $K\bar{K}$
intermediate state. In order to include this inelastic channel in the
IAM, the extended unitarity relation (\ref{eq:fefuni}) has to be
used above the $K\bar{K}$ threshold. This gives ${\rm Im}\Gamma
=-{\rm Im}F/|F|^2=-\sigma t/F-\sigma_{(K)}F^{\ast}_{(K)}
t_{(K)}/|F|^2$, which can be evaluated to two-loop order in the
chiral expansion
\begin{eqnarray}
\label{eq:finelaex}
{\rm Im}\Gamma (s) & = & -\sigma (s)\frac{t^I_l(s)}{F(s)}
-\sigma_{(K)}(s)\frac{F^{\ast}_{(K)}(s)t^I_{l(K)}(s)}{|F(s)|^2}
\nonumber \\
&\approx & -\sigma (s)\left\{ t^{I(0)}_l(s)\left[ 1-{\rm Re}
F^{(1)}(s)\right] +{\rm Re}t^{I(1)}_l(s)\right\} \nonumber \\
&&-\sigma_{(K)}(s)\left\{ t^{I(0)}_{l(K)}(s)\left[ F^{(0)}_{(K)}(s)
-2F^{(0)}_{(K)}(s){\rm Re}F^{(1)}(s)\right.\right. \nonumber \\
&&\left.\left. +{\rm Re}F^{(1)}_{(K)}(s)\right]
+F^{(0)}_{(K)}(s){\rm Re}t^{I(1)}_{l(K)}(s)\right\} .
\end{eqnarray}
This expression has to be used above the $K\bar{K}$ threshold in the
dispersion relation (\ref{eq:fdisp1}) for the IAM. In order to
simplify the relation between the IAM and ChPT in this inelastic case,
the perturbative extended unitarity relations (\ref{eq:fepuni}) may
be used to evaluate ${\rm Im}\Gamma^{(2)}$ to two-loop order in the
chiral expansion. Since this will give an expression exactly similar
to Eq. (\ref{eq:finelaex}), the inelastic IAM to two loops can also be
written in the form (\ref{eq:FPade2}) with the $K\bar{K}$ inelasticity
included in ChPT. In the same way, by expanding to one-loop order in
the chiral expansion, the inelastic IAM to one loop will be given by a
form similar to Eq. (\ref{eq:FPade1}). As was the case in the elastic
approximation, it is also expected that the inelastic higher order
corrections are substantially more suppressed for the IAM than for
ChPT. However, it still has to be investigated whether the IAM is
actually able to correctly describe the $K\bar{K}$ and other
inelasticities.

\subsection{Comparison with experiment}
\label{subsec:fcom}

The pion vector form factor is well known experimentally both in the
time-like region $s>4M^2_{\pi}$ \cite{ref:Bar85} and in the space-like
region $s<0$ ${\rm GeV}^2$ \cite{ref:Amen86,ref:Bebek78}. As for the
scalar form factor, this is not directly accessible to experiment.
However, it has been determined in terms of the experimental phase
shifts for the $\pi\pi /K\bar{K}$ system by a dispersive analysis
\cite{ref:GM90,ref:DGL90}. In the following, the theoretical form
factors will be limited to the elastic approximation when compared to
the experimental information.

\subsubsection{Vector form factor}
\label{subsubsec:vff}

The vector form factor to one loop depends on the single
scale-independent low-energy constant $\bar{l}_6$. This can be fixed
for both ChPT and the IAM from the experimental value of the pion
charge radius $\langle r^2\rangle_V=0.439\pm 0.008$ ${\rm fm}^2$
\cite{ref:Amen86}. With the central experimental value, one obtains the
results for $\bar{l}_6$ displayed in the first and third column in
Table \ref{Table3}, which is consistent with independent information
on this low-energy constant \cite{ref:Han96}. For two-loop ChPT, the
value of $\bar{l}_6$ should in principle be re-evaluated taking into
account the two-loop corrections. However, since a thorough
re-evaluation seems rather out of reach at the moment, the one-loop
value of $\bar{l}_6$ is also used in two-loop ChPT. Therefore, from
the central experimental value of the pion charge radius
\cite{ref:Amen86}, one obtains the result
that the two-loop contribution to
$\langle r^2\rangle_V$ contained in the coupling constant $\bar{f}_1$
vanishes. For the other coupling constant $\bar{f}_2$, this can be
fixed from the value $c_V=4.1$ ${\rm GeV}^{-4}$ \cite{ref:GM90} with a
result displayed in the second column in Table \ref{Table3}. Finally,
for the renormalized low-energy constants $l^r_1$, $l^r_2$, $l^r_3$,
and $l^r_4$ occuring in the $\pi\pi$ partial waves to one loop, they
will be given by the values shown in the second column in
Table \ref{Table1}.

For the IAM to two loops, the low-energy constant
$\bar{l}_6$ and the coupling constants $\bar{f}_1$ and $\bar{f}_2$ may
be determined simultaneously by fitting the time-like experimental
data below $\sqrt{s}=0.9$ GeV \cite{ref:Bar85} together with the
space-like experimental data \cite{ref:Amen86}. For the other
renormalized low-energy constants $l^r_1$, $l^r_2$, $l^r_3$, and
$l^r_4$, the values shown in the fourth column in Table \ref{Table1}
are used. This gives a rather good fit with values of $\bar{l}_6$,
$\bar{f}_1$, and $\bar{f}_2$ displayed in the fourth column in Table
\ref{Table3}. For the the pion charge radius $\langle r^2\rangle_V$
and the low-energy parameter $c_V$ also shown in Table \ref{Table3},
it is observed the the predictions for
the IAM to two loops agree remarkably well with the experimental
information $\langle r^2\rangle_V=0.439\pm 0.008$ ${\rm fm}^2$
\cite{ref:Amen86} and $c_V=4.1$ ${\rm GeV}^{-4}$ \cite{ref:GM90}. In
addition, with this method one finds that the two-loop correction to
$\langle r^2\rangle_V$ contained in the coupling constant $\bar{f}_1$
is approximately 20\% of the one-loop contribution, which is quite
reasonable. From Table \ref{Table3}, it is furthermore observed that
the prediction for $c_V$ obtained from the IAM to one loop also
agrees remarkably well with the experimental information, whereas the
corresponding prediction obtained from one-loop ChPT is far too small.

The vector form factor $|F_V|^2$ is shown for $-1$
${\rm GeV}^2\leq s\leq 1$
${\rm GeV}^2$ in Fig. \ref{Fig6}. The IAM to two loops describes the
data very well both in the time-like and in the space-like
regions. In fact, also the space-like data for $s<-0.26$ ${\rm GeV}^2$
\cite{ref:Bebek78}, which were not included in the fit, agree well
with the IAM to two loops. Of course, the small isospin violating
$\rho -\omega$ mixing is not reproduced by this method, since the
isospin limit has been applied throughout this paper. In addition,
this method should only be reliable below approximately $s=1$
${\rm GeV}^2$, since the elastic approximation has be used for the
vector form factor. As for the IAM to one loop, this agrees well with
the space-like experimental data. However, in the time-like region
this method only approximates the $\rho$(770) resonance, as has been
discussed in greater detail in Ref. \cite{ref:Han96}. Nevertheless, it
is obvious that the IAM to both one and two loops significantly
improves the behavior of ChPT, which only accounts for the low-energy
tail of the $\rho$(770) resonance. The agreement between one and
two-loop ChPT and the space-like experimental data \cite{ref:Amen86}
can be slightly improved by fitting these experimental data
\cite{ref:BC88}, but this will not change the range of applicability
of ChPT significantly.

In Fig. \ref{Fig7}, the predictions for the phase $\delta^1_1$ are
shown below $\sqrt{s}=1$ GeV. Only the IAM are
shown since one and two-loop ChPT will give the chiral $\pi\pi$ I=1 P
phase shifts to leading and one-loop order respectively. For the IAM
to two loops, the phase agrees very well with the experimental data
and gives a resonance at $M_{\rho}=0.771$ GeV defined to be at
$\delta^1_1(M_{\rho})=\pi /2$. Furthermore, since this method also
agrees well with the results of the IAM for the $\pi\pi$ I=1 P
phase shifts shown in Fig. \ref{Fig5}, the IAM to two loops satisfies
the elastic unitarity relation (\ref{eq:ffuni}) very well indeed. As
for the IAM to one loop, since this only approximates the experimental
data, the elastic unitarity relation (\ref{eq:ffuni}) will be somewhat
violated in this case. However, the IAM to one loop is also quite
consistent with a resonance structure \cite{ref:DT96} and will
give a resonance at $M_{\rho}=0.734$ GeV. Thus, the IAM to both one
and two loops are rather consistent approaches to extend the range of
applicability of ChPT in the case of the vector form factor.

\subsubsection{Scalar form factor}
\label{subsubsec:sff}

The normalized scalar form factor to one loop is given in terms of the
single scale-independent low-energy constant $\bar{l}_4$. The same
low-energy constant also occurs in $\pi\pi$ scattering to one loop,
where the central value $l^r_4(M_{\rho})=5.60\times 10^{-3}$ obtained
from the ratio $F_K/F_{\pi}$ \cite{ref:GL84,ref:GL85} was used.
Changing to scale-independent low-energy constants \cite{ref:GL84}
gives $\bar{l}_4=4.30$, which is used both in ChPT and the IAM to one
loop. The same value of $\bar{l}_4$ is also used in two-loop ChPT,
whereas the coupling constants $\bar{d}_1$ and $\bar{d}_2$ are
determined from the values $\langle r^2\rangle_S=0.60$ ${\rm fm}^2$
and $c_S=10.6$ ${\rm GeV}^{-4}$ \cite{ref:GM90,ref:DGL90},
respectively. For the additional low-energy constants to two loops
$l^r_1$, $l^r_2$, and $l^r_3$, they will be given by the values shown
in the second column in Table \ref{Table1}.

For the IAM to two
loops, the low-energy constant $\bar{l}_4$ and the coupling constants
$\bar{d}_1$ and $\bar{d}_2$ cannot be determined in the same way
as was applied for the vector form factor. This is because the scalar
form factor is not directly accessible to experiment. However, in the
elastic region the phase of the scalar form factor is given by the
$\pi\pi$ I=0 S phase shift, which is known experimentally. Therefore,
the coupling constants $\bar{d}_1$ and $\bar{d}_2$ may be determined
by fitting the experimental phase shift $\delta^0_0$ below
$\sqrt{s}=0.9$ GeV \cite{ref:Ros77,ref:Proto73,ref:Hyams73,ref:EM74},
whereas $\bar{l}_4$ can be fixed from the requirement
$\langle r^2\rangle_S=0.60$ ${\rm fm}^2$. With the remaining
low-energy constants $l^r_1$, $l^r_2$, and $l^r_3$ given by the values
in the fourth column in Table \ref{Table1}, this gives an acceptable
fit for the $\pi\pi$ I=0 S phase shift.

In Table \ref{Table4}, the results for the low-energy constant
$\bar{l}_4$ and the coupling constants $\bar{d}_1$ and $\bar{d}_2$ are
shown both for ChPT and the IAM. As for the IAM to two loops, the value
of $\bar{l}_4$ is slightly different from the one-loop value
$\bar{l}_4=4.30$. This might well be due to the inclusion of the
higher order corrections in the determination of this low-energy
constant in the two-loop case. However, the value of $\bar{l}_4$ for
the IAM to two loops is still consistent with the value
$\bar{l}_4=4.3\pm 0.9$ obtained from the ratio $F_K/F_{\pi}$ to
one-loop order in ChPT \cite{ref:GL84,ref:GL85}. In fact, using the
value $\bar{l}_4=3.68$ in the IAM to two loops, the previous numerical
analysis of $\pi\pi$ scattering and the pion vector form factor will
only change insignificantly. The largest numerical difference will be
for the $\pi\pi$ threshold parameters, but the values given for the
IAM to two loops in Table \ref{Table2} will at most change by 5\%. In
the future it might be possible to determine the two-loop value of
$\bar{l}_4$ from independent observables.

The corresponding values of the pion scalar radius
$\langle r^2\rangle_S$ and the low-energy parameter $c_S$ are
also given in Table \ref{Table4}. For the pion scalar radius,
the predictions obtained from
ChPT and the IAM to one loop agree rather well with the result from
the dispersive analysis $\langle r^2\rangle_S=0.60$ ${\rm fm}^2$
(solution B) \cite{ref:GM90,ref:DGL90}. For the low-energy parameter
$c_S$, the predictions from the IAM to one and two loops agree within
10-15\% with the result $c_S=10.6$ ${\rm GeV}^{-4}$ (solution B)
\cite{ref:GM90,ref:DGL90}, whereas the corresponding prediction from
one-loop ChPT is too small.

The scalar form factor is known up to $\sqrt{s}=0.7$ GeV from the
dispersive analysis in Refs. \cite{ref:GM90,ref:DGL90}. In Fig.
\ref{Fig8}, solution B from this dispersive analysis is compared to the
scalar form factor obtained from the IAM and ChPT. It is observed that
the IAM to two loops agrees very well with the dispersive analysis
over the whole energy region, whereas the same is only true for the
IAM to one loop below approximately 0.5 GeV. As for two-loop ChPT,
this works rather well up to about 0.4 GeV, while one-loop ChPT only
agrees with the dispersive analysis below the $\pi\pi$ threshold. This
is due to the strong final-state interaction, which makes the higher
order corrections important even at low energies.

In Fig. \ref{Fig9}, the phase $\delta^0_0$ obtained from the IAM is
shown below $\sqrt{s}=1$ GeV. For the IAM to two loops, it is observed
that below approximately 0.9 GeV, the phase agrees rather well with
both the experimental data and the two-loop IAM result for the
$\pi\pi$ I=0 S phase shift shown in Fig. \ref{Fig3}. Thus, the IAM to
two loops satisfies the elastic unitarity relation (\ref{eq:ffuni})
rather well. Slightly above 0.9 GeV, the IAM to two loops shows a very
rapid phase variation of approximately $180^{\circ}$, which might be
a signal indicating that this method could reproduce the $f_0$(980)
resonance. However, in order to investigate this the $K\bar{K}$
inelasticity must be included in the IAM. As for the IAM to one loop,
this method only describes the data well at rather low energies. The
agreement could be improved at somewhat higher energies by reducing
the value of the low-energy constant $\bar{l}_4$ within the error bars
$\bar{l}_4=4.3\pm 0.9$, but this will at the same time give a value
for the pion scalar radius less consistent with the dispersive result
$\langle r^2\rangle_S=0.60$ ${\rm fm}^2$.

\begin{figure}
\centering{\epsfig{file=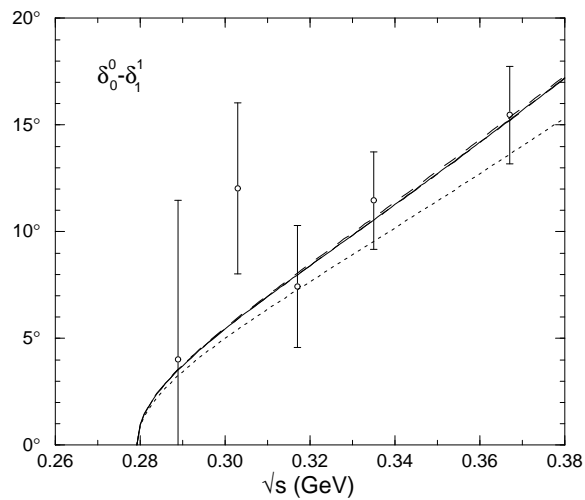,angle=-90}}
\caption{The phase shift difference $\delta^0_0-\delta^1_1$ below
$\protect\sqrt{s}=0.38$ GeV. The solid line is the IAM to two loops,
the dashed line the IAM to one loop, the dashed-dotted line two-loop
ChPT, and the dotted line one-loop ChPT. The solid, dashed, and
dashed-dotted curves are very close to each other and therefore hard
to distinguish from each other in the figure. The experimental data
are from Ref. \protect\cite{ref:Ros77}.}
\label{Fig2}
\end{figure}

\begin{figure}
\centering{\epsfig{file=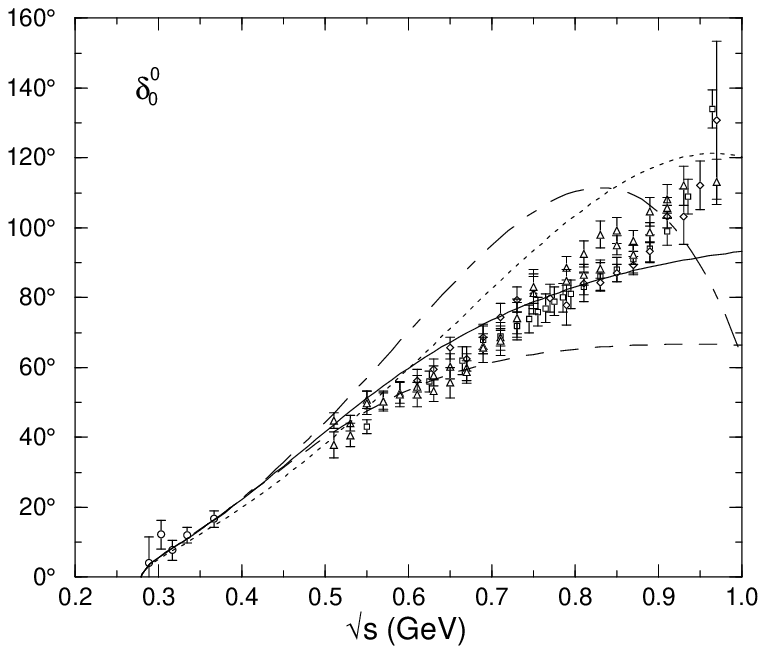,angle=-90}}
\caption{The phase shift $\delta^0_0$ below $\protect\sqrt{s}=1$ GeV.
The curves are as in Fig. \protect\ref{Fig2} and the experimental data
are from Ref. \protect\cite{ref:Ros77} (circles), Ref.
\protect\cite{ref:Proto73} (squares), Ref. \protect\cite{ref:Hyams73}
(diamonds), and Ref. \protect\cite{ref:EM74} (triangles).}
\label{Fig3}
\end{figure}

\begin{figure}
\centering{\epsfig{file=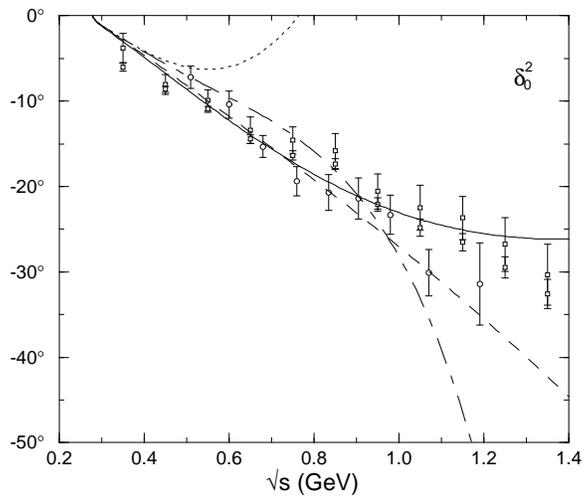,angle=-90}}
\caption{The phase shift $\delta^2_0$ below $\protect\sqrt{s}=1.4$
GeV. The curves are as in Fig. \protect\ref{Fig2} and the
experimental data are from Ref. \protect\cite{ref:Losty74} (circles)
and Ref. \protect\cite{ref:Hoog77} (squares).}
\label{Fig4}
\end{figure}

\begin{figure}
\centering{\epsfig{file=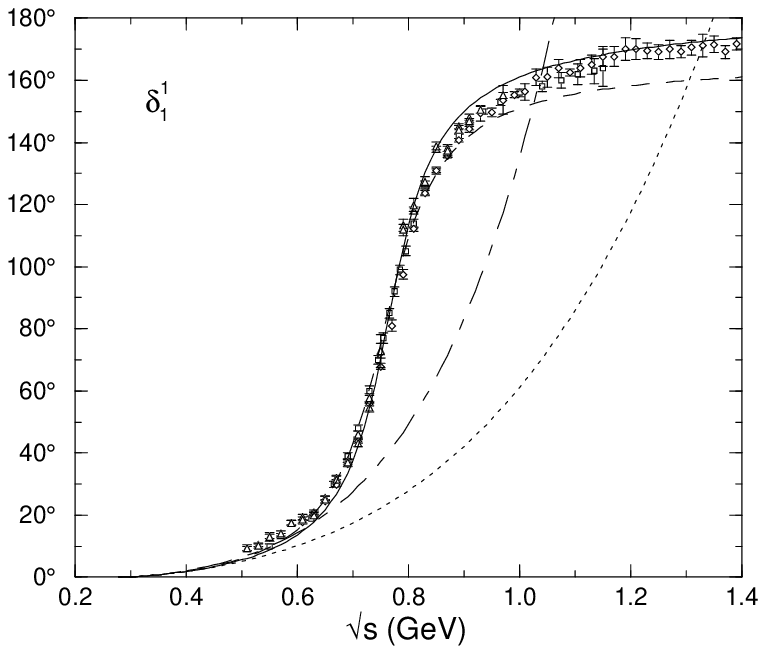,angle=-90}}
\caption{The phase shift $\delta^1_1$ below $\protect\sqrt{s}=1.4$
GeV. The curves are as in Fig. \protect\ref{Fig2} and the
experimental data are from Ref. \protect\cite{ref:Proto73} (squares),
Ref. \protect\cite{ref:Hyams73}
(diamonds), and Ref. \protect\cite{ref:EM74} (triangles).}
\label{Fig5}
\end{figure}

\begin{figure}
\centering{\epsfig{file=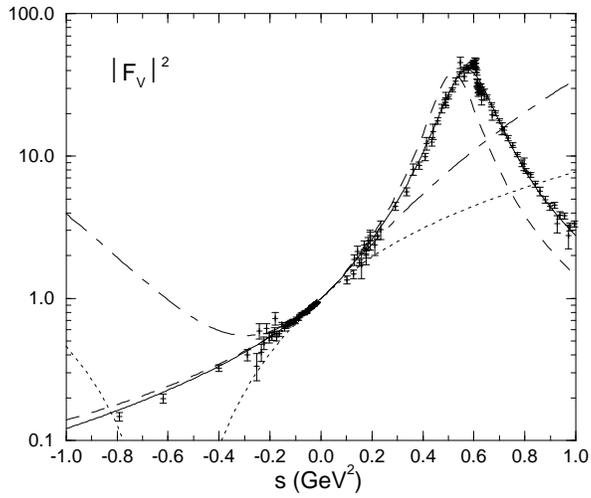,angle=-90}}
\caption{The vector form factor $|F_V|^2$ for $-1$
${\rm GeV}^2\leq s\leq 1$
${\rm GeV}^2$. The curves are as in Fig. \protect\ref{Fig2} and the
experimental data are from Ref. \protect\cite{ref:Bar85}, Ref.
\protect\cite{ref:Amen86}, and Ref. \protect\cite{ref:Bebek78}.}
\label{Fig6}
\end{figure}

\begin{figure}
\centering{\epsfig{file=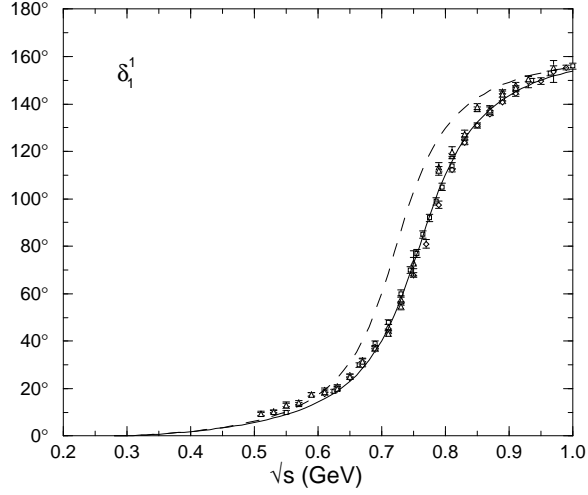,angle=-90}}
\caption{The phase $\delta^1_1$ for the vector form factor below
$\protect\sqrt{s}=1$ GeV. The solid line is the IAM to two loops, the
dashed line the IAM to one loop, and the experimental data are as in
Fig. \protect\ref{Fig5}.}
\label{Fig7}
\end{figure}

\begin{figure}
\centering{\epsfig{file=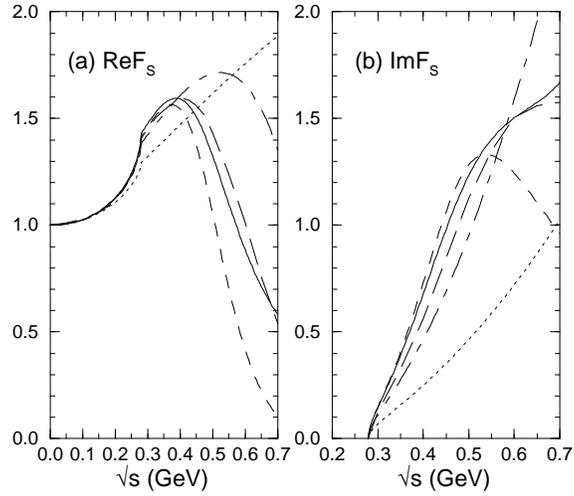,angle=-90}}
\caption{The real (a) and imaginary (b) part of the scalar form factor
$F_S$ below $\protect\sqrt{s}=0.7$ GeV. The long-dashed curve is
solution B from a dispersive analysis
\protect\cite{ref:GM90,ref:DGL90}, whereas the other curves are as in
Fig. \protect\ref{Fig2}.}
\label{Fig8}
\end{figure}

\begin{figure}
\centering{\epsfig{file=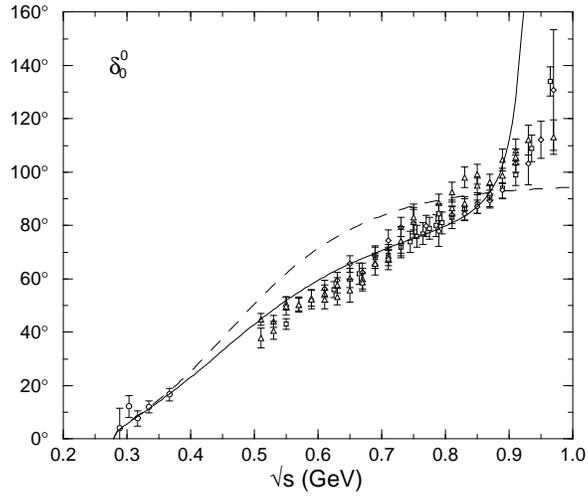,angle=-90}}
\caption{The phase $\delta^0_0$ for the scalar form factor below
$\protect\sqrt{s}=1$ GeV. The curves are as in Fig. \protect\ref{Fig7}
and the experimental data are as in Fig. \protect\ref{Fig3}.}
\label{Fig9}
\end{figure}

\begin{figure}
\centering{\epsfig{file=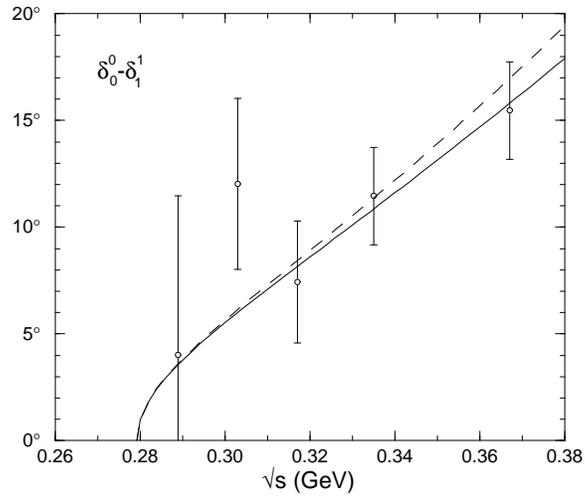,angle=-90}}
\caption{The phase difference $\delta^0_0-\delta^1_1$ between the
scalar and vector form factors below $\protect\sqrt{s}=0.38$ GeV. The
curves are as in Fig. \protect\ref{Fig7} and the experimental data are
as in Fig. \protect\ref{Fig2}.}
\label{Fig10}
\end{figure}

Finally, the phase difference $\delta^0_0-\delta^1_1$ between the
scalar and vector form factors is shown in Fig. \ref{Fig10} below
$\sqrt{s}=0.38$ GeV. The IAM to two loops agrees very well with both
the central experimental data and the results of the IAM for the
$\pi\pi$ phase shift difference shown in Fig. \ref{Fig2}. The IAM to
one loop also agrees with the experimental data within the error bars,
but is slightly different from the corresponding phase shift
differences shown in Fig. \ref{Fig2}. As already pointed out, it is
expected that forthcoming kaon facilities like DA$\Phi$NE can
reduce the error bars on $\delta^0_0-\delta^1_1$ significantly
\cite{ref:BF95}, which will allow one to test the
consistency of the IAM at low energies in greater detail.

\section{CONCLUSIONS}
\label{sec:con}

The IAM is based upon the use of unitarity and dispersion relations
together with the chiral expansion. It has previously been shown that
this method applied to one-loop order successfully extends the range of
applicability of ChPT. However, the IAM is not restricted to the
one-loop approximation but can in general be applied to any given
order in the chiral expansion. In the present paper, the extension of
the IAM to two loops has been analysed in the case of $\pi\pi$
scattering and the pion form factors. The analysis has mainly been
restricted to the elastic approximation but the possible extension to
the inelastic case has also been discussed in some detail.

From the derivation of the IAM applied to two loops in the chiral
expansion, it is found that this approach indeed improves
the IAM to one loop both for $\pi\pi$ scattering and in
the case of the pion form factors. For $\pi\pi$ scattering, a
comparison with the experimental data shows that the IAM applied to
two loops is in fact rather consistent with the experimental data up
to energies where inelasticities become essential. At somewhat lower
energies, the two-loop approach compares well with the one-loop
approximation, and in
the threshold region they both agree with ChPT. In the case of the
pion form factors, it is also found that the IAM to two loops is
consistent with the experimental information up to the first important
inelastic threshold. For the IAM to one loop, although this also
successfully extends the range of applicability of ChPT, it only
approximates the form factors at higher energies.

Thus, the present analysis suggests that the IAM is indeed a rather
systematic way of extending the range of applicability of ChPT.
However, in order to investigate this in greater detail, the IAM to
two loops should be applied to other processes. In this way, some of
the two-loop low-energy constants in the IAM might be determined from
independent observables. There are also other ways of extending the
range of applicability of ChPT and work is still in progress in this
subject. However, at present the IAM provides one of the most
intriguing methods to improve ChPT in a systematic manner.

\acknowledgments

The author thanks A. Miranda and G. C. Oades for discussions and
comments, and J. Bijnens for a private communication regarding the
two-loop low-energy constants estimated from resonance exchange. The
support from The Faculty of Science, Aarhus University is also
acknowledged.

\begin{table}
\caption{Values of the low-energy constants at the $\rho$(770) scale
used for one-loop ChPT (ChPT1), two-loop ChPT (ChPT2), the IAM to one
loop (IAM1), and the IAM to two loops (IAM2). For a discussion on how
these low-energy constants have been determined see the text.}
\label{Table1}
\begin{tabular}{ccccc}
&ChPT1&ChPT2&IAM1&IAM2\\
\tableline
$10^3l^r_1(M_{\rho})$&-5.40&-5.40&-4.13&-3.57\\
$10^3l^r_2(M_{\rho})$&5.67&5.67&4.05&1.21\\
$10^3l^r_3(M_{\rho})$&0.82&0.82&0.82&0.82\\
$10^3l^r_4(M_{\rho})$&5.60&5.60&5.60&5.60\\
$10^5r^r_1(M_{\rho})$&&-6.1&&-6.1\\
$10^5r^r_2(M_{\rho})$&&13.0&&13.0\\
$10^5r^r_3(M_{\rho})$&&-17.0&&-17.0\\
$10^5r^r_4(M_{\rho})$&&-10.1&&-10.1\\
$10^5r^r_5(M_{\rho})$&&11.4&&16.9\\
$10^5r^r_6(M_{\rho})$&&3.0&&-6.5\\
\end{tabular}
\end{table}

\begin{table}
\caption{Threshold parameters obtained from current algebra (CA),
one-loop ChPT (ChPT1), two-loop ChPT (ChPT2), the IAM to one loop
(IAM1), and the IAM to two loops (IAM2). The experimental data are
from Ref. \protect\cite{ref:Nagels79} except for $b^1_1$, which is
from Ref. \protect\cite{ref:ATW96}.}
\label{Table2}
\begin{tabular}{ccccccc}
&CA&ChPT1&ChPT2&IAM1&IAM2&Experiment\\
\tableline
$2a^0_0-5a^2_0$&0.545&0.623&0.654&0.659&0.665&0.657$\pm$0.052\\
$a^0_0$&0.159&0.205&0.222&0.222&0.221&0.26$\pm$0.05\\
$b^0_0$&0.182&0.254&0.282&0.283&0.278&0.25$\pm$0.03\\
-$10a^2_0$&0.454&0.425&0.420&0.429&0.447&0.28$\pm$0.12\\
-$10b^2_0$&0.908&0.736&0.727&0.768&0.809&0.82$\pm$0.08\\
$10a^1_1$&0.303&0.378&0.405&0.383&0.375&0.38$\pm$0.02\\
$10^2b^1_1$&&0.502&0.830&0.622&0.494&0.6$\pm$0.4\\
\end{tabular}
\end{table}

\begin{table}
\caption{Values of the scale-independent low-energy constant
$\bar{l}_6$ and the coupling constants $\bar{f}_1$ and $\bar{f}_2$
together with the corresponding values of the pion charge radius
$\langle r^2\rangle_V$ and the low-energy parameter $c_V$.
The notation is as in Table \protect\ref{Table1}.}
\label{Table3}
\begin{tabular}{ccccc}
&ChPT1&ChPT2&IAM1&IAM2\\
\tableline
$\bar{l}_6$&16.20&16.20&16.20&13.19\\
$\bar{f}_1$&&0.0&&197.3\\
$\bar{f}_2$&&6.3&&3.7\\
$\langle r^2\rangle_V$ (${\rm fm}^2$)&0.439&0.439&0.439&0.434\\
$c_V$ (${\rm GeV}^{-4}$)&0.6&4.1&4.2&3.9\\
\end{tabular}
\end{table}

\begin{table}
\caption{Values of the scale-independent low-energy constant
$\bar{l}_4$ and the coupling constants $\bar{d}_1$ and $\bar{d}_2$
together with the corresponding values of the pion scalar radius
$\langle r^2\rangle_S$ and the low-energy parameter $c_S$.
The notation is as in Table \protect\ref{Table1}.}
\label{Table4}
\begin{tabular}{ccccc}
&ChPT1&ChPT2&IAM1&IAM2\\
\tableline
$\bar{l}_4$&4.30&4.30&4.30&3.68\\
$\bar{d}_1$&&17.0&&59.8\\
$\bar{d}_2$&&8.3&&6.0\\
$\langle r^2\rangle_S$ (${\rm fm}^2$)&0.56&0.60&0.56&0.60\\
$c_S$ (${\rm GeV}^{-4}$)&6.0&10.6&11.7&12.2\\
\end{tabular}
\end{table}

\end{document}